\documentclass[12pt]{iopart}

\usepackage{color} 
\usepackage{graphicx}
\usepackage{hyperref}

\begin{document}

\title[First cryogenic operation of KAGRA]{First cryogenic test operation of underground km-scale gravitational-wave observatory KAGRA}

%--------------------------------------------------------------------------
% Author List
%--------------------------------------------------------------------------
\author{
T.~Akutsu$^{1, 2}$, 
M.~Ando$^{3, 4, 1}$, 
K.~Arai$^{5}$, 
Y.~Arai$^{5}$, 
S.~Araki$^{6}$, 
A.~Araya$^{7}$, 
N.~Aritomi$^{3}$, 
H.~Asada$^{8}$, 
Y.~Aso$^{9, 10}$, 
S.~Atsuta$^{11}$
% ...
\\
(KAGRA collaboration)
}
\address{${}^{1}$ National Astronomical Observatory of Japan (NAOJ), Mitaka City, Tokyo 181-8588, Japan }
\address{${}^{2}$ Advanced Technology Center, National Astronomical Observatory of Japan (NAOJ), Mitaka City, Tokyo 181-8588, Japan }
\address{${}^{3}$ Department of Physics, The University of Tokyo, Bunkyo-ku, Tokyo 113-0033, Japan }
\address{${}^{4}$ Research Center for the Early Universe (RESCEU), The University of Tokyo, Bunkyo-ku, Tokyo 113-0033, Japan }
\address{${}^{5}$ Institute for Cosmic Ray Research (ICRR), KAGRA Observatory, The University of Tokyo, Kashiwa City, Chiba 277-8582, Japan }
\address{${}^{6}$ Accelerator Laboratory, High Energy Accelerator Research Organization (KEK), Tsukuba City, Ibaraki 305-0801, Japan }
\address{${}^{7}$ Earthquake Research Institute, The University of Tokyo, Bunkyo-ku, Tokyo 113-0032, Japan }
\address{${}^{8}$ Department of Mathematics and Physics, Hirosaki University, Hirosaki City, Aomori 036-8561, Japan }
\address{${}^{9}$ Kamioka Branch, National Astronomical Observatory of Japan (NAOJ), Kamioka-cho, Hida City, Gifu 506-1205, Japan }
\address{${}^{10}$ The Graduate University for Advanced Studies (SOKENDAI), Mitaka City, Tokyo 181-8588, Japan }
\address{${}^{11}$ Graduate School of Science and Technology, Tokyo Institute of Technology, Meguro-ku, Tokyo 152-8551, Japan }
\address{${}^{12}$ Institute for Cosmic Ray Research (ICRR), KAGRA Observatory, The University of Tokyo, Kamioka-cho, Hida City, Gifu 506-1205, Japan }
\address{${}^{13}$ Korea Institute of Science and Technology Information (KISTI), Yuseong-gu, Daejeon 34141, Korea }
\address{${}^{14}$ Graduate School of Science, Osaka University, Toyonaka City, Osaka 560-0043, Japan }
\address{${}^{15}$ Department of Physics, National Taiwan Normal University, sec. 4, Taipei 116, Taiwan, R.O.C. }
\address{${}^{16}$ Department of Physics, Sogang University, Mapo-Gu, Seoul 121-742, Korea }
\address{${}^{17}$ Institute of Physics, Academia Sinica, Nankang, Taipei 11529, Taiwan, R.O.C. }
\address{${}^{18}$ Department of Physics, University of Toyama, Toyama City, Toyama 930-8555, Japan }
\address{${}^{19}$ Univ. Grenoble Alpes, Laboratoire d'Annecy de Physique des Particules (LAPP), Universit\'e Savoie Mont Blanc, CNRS/IN2P3, F-74941 Annecy, France }
\address{${}^{20}$ Department of Astronomy, The University of Tokyo, Bunkyo-ku, Tokyo 113-0033, Japan }
\address{${}^{21}$ Applied Research Laboratory, High Energy Accelerator Research Organization (KEK), Tsukuba City, Ibaraki 305-0801, Japan }
\address{${}^{22}$ Department of Applied Physics, Fukuoka University, Jonan, Fukuoka City, Fukuoka 814-0180, Japan }
\address{${}^{23}$ Faculty of Engineering, University of Toyama, Toyama City, Toyama 930-8555, Japan }
\address{${}^{24}$ Institute for Cosmic Ray Research (ICRR), Research Center for Cosmic Neutrinos (RCCN), The University of Tokyo, Kashiwa City, Chiba 277-8582, Japan }
\address{${}^{25}$ Department of Physics , National Central University, Zhongli District, Taoyuan City 32001, Taiwan, R.O.C. }
\address{${}^{26}$ Center for High Energy and High Field Physics, Zhongli District, Taoyuan City 32001, Taiwan, R.O.C. }
\address{${}^{27}$ Yukawa Institute for Theoretical Physics (YITP), Kyoto University, Sakyou-ku, Kyoto City, Kyoto 606-8502, Japan }
\address{${}^{28}$ Graduate School of Science, Osaka City University, Sumiyoshi-ku, Osaka City, Osaka 558-8585, Japan }
\address{${}^{29}$ JAXA Institute of Space and Astronautical Science, Chuo-ku, Sagamihara City, Kanagawa 252-0222, Japan }
\address{${}^{30}$ Institute for Cosmic Ray Research (ICRR), The University of Tokyo, Kashiwa City, Chiba 277-8582, Japan }
\address{${}^{31}$ Department of Physics, Ewha Womans University, Seodaemun-gu, Seoul 03760, Korea }
\address{${}^{32}$ Department of Physics, Myongji University, Yongin 449-728, Korea }
\address{${}^{33}$ Department of Computer Simulation, Inje University, Gimhae, Gyeongsangnam-do 50834, Korea }
\address{${}^{34}$ National Institute for Mathematical Sciences, Daejeon 34047, Korea }
\address{${}^{35}$ School of Natural Science, Ulsan National Institute of Science and Technology (UNIST), Ulsan 44919, Korea }
\address{${}^{36}$ Department of Physical Science, Hiroshima University, Higashihiroshima City, Hiroshima 903-0213, Japan }
\address{${}^{37}$ Department of Physics and Institute of Astronomy, National Tsing Hua University, Hsinchu 30013, Taiwan, R.O.C. }
\address{${}^{38}$ Institute for Cosmic Ray Research (ICRR), Research Center for Cosmic Neutrinos (RCCN), The University of Tokyo, Kamioka-cho, Hida City, Gifu 506-1205, Japan }
\address{${}^{39}$ California Institute of Technology, Pasadena, CA 91125, USA }
\address{${}^{40}$ Institute for Advanced Research, Nagoya University, Furocho, Chikusa-ku, Nagoya City, Aichi 464-8602, Japan }
\address{${}^{41}$ Department of Physics, Hanyang University, Seoul 133-791, Korea }
\address{${}^{42}$ Korea Astronomy and Space Science Institute (KASI), Yuseong-gu, Daejeon 34055, Korea }
\address{${}^{43}$ National Center for High-performance computing, National Applied Research Laboratories, Hsinchu Science Park, Hsinchu City 30076, Taiwan, R.O.C. }
\address{${}^{44}$ Department of Physics, Tamkang University, Danshui Dist., New Taipei City 25137, Taiwan, R.O.C. }
\address{${}^{45}$ Department of Advanced Materials Science, The University of Tokyo, Kashiwa City, Chiba 277-8582, Japan }
\address{${}^{46}$ Istituto Nazionale di Fisica Nucleare, Sapienza University, Roma 00185, Italy }
\address{${}^{47}$ Department of Mathematical Analysis and Statistical Inference, The Institute of Statistical Mathematics, Tachikawa City, Tokyo 190-8562, Japan }
\address{${}^{48}$ School of Physics, Korea Institute for Advanced Study (KIAS) , Seoul 02455, Korea }
\address{${}^{49}$ Institute for Photon Science and Technology, The University of Tokyo, Bunkyo-ku, Tokyo 113-8656, Japan }
\address{${}^{50}$ Disaster Prevention Research Institute, Kyoto University, Uji City, Kyoto 611-0011, Japan }
\address{${}^{51}$ Institute for Laser Science, University of Electro-Communications, Chofu City, Tokyo 182-8585, Japan }
\address{${}^{52}$ The Applied Electromagnetic Research Institute, National Institute of Information and Communications Technology (NICT), Koganei City, Tokyo 184-8795, Japan }
\address{${}^{53}$ Department of Physics, Kyoto University, Sakyou-ku, Kyoto City, Kyoto 606-8502, Japan }
\address{${}^{54}$ Faculty of Law, Ryukoku University, Fushimi-ku, Kyoto City, Kyoto 612-8577, Japan }
\address{${}^{55}$ Department of Physics , University of Notre Dame, Notre Dame, IN 46556, USA }
\address{${}^{56}$ Department of Physics, Wuhan Institute of Physics and Mathematics, CAS, Xiaohongshan, Wuhan 430071, China }
\address{${}^{57}$ School of Optical Electrical and Computer Engineering, The University of Shanghai for Science and Technology, Shanghai 200093, China }
\address{${}^{58}$ Faculty of Engineering, Niigata University, Nishi-ku, Niigata City, Niigata 950-2181, Japan }
\address{${}^{59}$ Graduate School of Science and Technology, Niigata University, Nishi-ku, Niigata City, Niigata 950-2181, Japan }
\address{${}^{60}$ Center for Measurement Standards, Industrial Technology Research Institute, Hsinchu, 30011, Taiwan, R.O.C. }
\address{${}^{61}$ Department of Engineering, University of Sannio, Benevento 82100, Italy }
\address{${}^{62}$ Faculty of Arts and Science, Kyushu University, Nishi-ku, Fukuoka City, Fukuoka 819-0395, Japan }
\address{${}^{63}$ Research Institute for Science and Engineering, Waseda University, Shinjuku, Tokyo 169-8555, Japan }
\address{${}^{64}$ Department of Electronic Control Engineering, National Institute of Technology, Nagaoka College, Nagaoka City, Niigata 940-8532, Japan }
\address{${}^{65}$ Kavli Institute for the Physics and Mathematics of the Universe (IPMU), Kashiwa City, Chiba 277-8583, Japan }
\address{${}^{66}$ Department of Information \& Management Systems Engineering, Nagaoka University of Technology, Nagaoka City, Niigata 940-2188, Japan }
\address{${}^{67}$ Graduate School of Science and Engineering, Hosei University, Koganei City, Tokyo 184-8584, Japan }
\address{${}^{68}$ Faculty of Science, Toho University, Funabashi City, Chiba 274-8510, Japan }
\address{${}^{69}$ Faculty of Information Science and Technology, Osaka Institute of Technology, Hirakata City, Osaka 573-0196, Japan }
\address{${}^{70}$ School of High Energy Accelerator Science, The Graduate University for Advanced Studies (SOKENDAI), Tsukuba City, Ibaraki 305-0801, Japan }
\address{${}^{71}$ Istituto Nazionale di Fisica Nucleare, University of Perugia, Perugia 06123, Italy }
\address{${}^{72}$ Faculty of Science, Niigata University, Nishi-ku, Niigata City, Niigata 950-2181, Japan }
\address{${}^{73}$ Department of Communications, National Defense Academy of Japan, Yokosuka City, Kanagawa 239-8686, Japan }
\address{${}^{74}$ Department of Physics, University of Florida, Gainesville, FL 32611, USA  }
\address{${}^{75}$ Department of Physics and Astronomy, Sejong University, Gwangjin-gu, Seoul 143-747, Korea }
\address{${}^{76}$ Department of Physics, Korea University, Seongbuk-gu, Seoul 02841, Korea }
\address{${}^{77}$ Department of Astronomy, Beijing Normal University, Beijing 100875, China }

\ead{Sadakazu Haino Sadakazu.Haino@cern.ch, 
Nobuyuki Kanda kanda@sci.osaka-cu.ac.jp, 
Yuta Michimura michimura@granite.phys.s.u-tokyo.ac.jp, 
Hisaaki Shinkai hisaaki.shinkai@oit.ac.jp, 
Takahiro Yamamoto yamat@icrr.u-tokyo.ac.jp}
\vspace{10pt}
\begin{indented}
\item[] (\today)
\end{indented}

%--------------------------------------------------------------------------
% Abstract
%--------------------------------------------------------------------------
\begin{abstract}
KAGRA is a second-generation interferometric gravitational-wave detector with 3-km arms constructed at Kamioka, Gifu in Japan.  It is now in its final installation phase, which we call {\it bKAGRA} (baseline KAGRA), with scientific observations expected to begin in late 2019.
One of the advantages of KAGRA is its underground location of at least 200~m below the ground surface,  which brings small seismic motion at low frequencies and high stability of the detector.
Another advantage is that it cools down the sapphire test mass mirrors to cryogenic temperatures to reduce thermal noise.
In April--May 2018, we have operated a 3-km Michelson interferometer with a cryogenic test mass for 10 days, which was the first time that km-scale interferometer was operated at cryogenic temperatures. In this article, we report the results of this ``bKAGRA Phase 1" operation.
We have demonstrated the feasibility of 3-km interferometer alignment and control with cryogenic mirrors.
\end{abstract}

%--------------------------------------------------------------------------
% Section 1 
%--------------------------------------------------------------------------
\section{Introduction}
Direct detection of gravitational waves (GWs) in the past three years have proved that we are entering a new era of physics and astronomy. LIGO Scientific Collaboration and Virgo Collaboration have so far reported ten binary black hole mergers (GW150914~\cite{GW150914PRL}, and else~\cite{GWTC1Catalog}), and one binary neutron star marger (GW170817~\cite{GW170817PRL}) during their first and second observing runs (O1 \& O2).  According to the latest release~\cite{GWTC1Catalog}, the sources of GWs vary in distance from $320^{+120}_{-110}$~Mpc to $2750^{+1350}_{-1320}$~Mpc and total-mass for the black hole binaries range from $18.6^{+3.1}_{-0.7}M_\odot$ to  $85.1^{+15.6}_{-10.9}M_\odot$. LIGO and Virgo also list fourteen additional marginal events~\cite{GWTC1Catalogweb}.
These numbers tell us that hunting for GWs is now a part of astronomy, that is, we now have a totally new set of ``eyes" to observe the Universe.

Information from GWs enables us not only to identify the parameters of compact objects, but to also model and provide constraints of many physical and astrophysical phenomena. 
As for the neutron-star binary merger (GW170817), after the short notice of the detection by LIGO and Virgo, successive observations of the source using $\gamma$-ray, X-ray, UV, optical, IR, and radio telescopes were made. These observations identified the source, and revealed the nature of the gigantic astrophysical event; details such as the plausible evidence of a rapid process of nuclear fusion, constraints to equation of state of nuclear matter, constraints to cosmological models, and so on.  In the near future, we are quite certain that strong gravitational effects from black holes will provide us a chance to test Einstein's general relativity. 
Accumulation of detections will also allow us to discuss the formation processes of stars and binaries. 

The LIGO Scientific Collaboration has two 4-km arm laser interferometers at Hanford, Washington state, and at Livingston, Louisiana state. The independent detection from these two detectors provided the first 5 detections of GWs.  The Virgo Collaboration has a 3-km arm laser interferometer at Pisa, Italy, and its joint observation with LIGO in August 2017 resulted in detection with high precision in its source localization, which helped a lot in motivating the follow-up observation for astronomers.  In principle, GWs should be detected independently at different locations in order to rule out false positives.  If three detectors are operating simultaneously, we are able to identify the location of the source.  With more detectors, we will get more precise information of parameters, including the polarization of GWs.  With this in mind, KAGRA is the next interferometer joining to the GW network. 

KAGRA is a 3-km laser interferometer, constructed in Kamioka, Gifu, Japan, and is now in its final installation phase~\cite{SomiyaKAGRA,AsoKAGRA}\footnote{KAGRA was originally called LCGT (large-scale cryogenic gravitational-wave telescope), but was renamed after the approval of the project via a public naming contest.  The name KAGRA remind us of Kamioka (the location) plus gravity, and is also the Japanese word for traditional sacred music and dance in front of the gods. 
A detailed history of KAGRA is described in \cite{NatureAstronomy201811}.}.
KAGRA is operated at cryogenic temperatures to reduce thermal noise around the detector's most sensitive band at around 100~Hz, and located at underground for smaller seismic noise. The future GW detector concepts such as Einstein Telescope~\cite{ET} and Cosmic Explorer~\cite{CE} consider to use these techniques, and therefore demonstrating the techniques with KAGRA is a key to the next generation GW detectors.

Since KAGRA is located far from both LIGO and Virgo, KAGRA's operation is expected to make the source localization and waveform reconstruction more precise, when all three observatories (LIGO, Virgo and KAGRA) perform the observation jointly~\cite{KLV:ScenarioPaper}. Moreover, increasing the number of detectors is intrinsically essential to search for non-tensorial polarization modes of GWs to test general relatvity~\cite{TakedaPol,HagiwaraPol}.

The KAGRA project is split into two stages, initial KAGRA ({\it iKAGRA}) and baseline KAGRA ({\it bKAGRA}), as summarized in Table~\ref{table:ibKAGRA}. In the iKAGRA stage, we had constructed the basic infrastructure including the tunnel and vacuum systems, and formed a simple 3-km Michelson interferometer that consisted of two end test masses and a beam splitter. The mirrors were fused silica mirrors at room temperature suspended by simplified systems. We operated iKAGRA in March and April of 2016, which was the first kilometer-scale interferometer operated underground.  We first had to overcome challenges such as measuring the exact positions of the vacuum systems and the synchronization of digital real-time control systems 3 km apart in the tunnel using the Global Positioning System. In the iKAGRA operation, we demonstrated that the vacuum tubes are aligned well enough to form the interferometer, and real-time interferometer control was possible. For details of the iKAGRA operation, see Ref.~\cite{iKAGRA}.

The bKAGRA stage is focused on GW observations with a cryogenic resonant side-band extraction (RSE) interferometer. Before operating a full interferometer, we first constructed a 3-km Michelson interferometer with two sapphire test masses. All the mirrors were suspended by full systems, and one of the test masses was cooled down to cryogenic temperatures. This phase is called {\it bKAGRA Phase 1}, and we operated the interferometer from April 28 to May 6 in 2018.  This operation was the first cryogenic operation with full suspension systems.  The purpose of this article is to report the results of this operation.  We characterized the suspensions and tested the compatibility of cryogenic cooling with 3-km interferometer alignment and control. 

\begin{table}
  \begin{center}
    \caption{Summary of the phases of KAGRA. Type-C suspension is a double pendulum and Type-A suspension is a full eight-stage pendulum (see Section 2 for details). RSE: resonant sideband extraction interferometer.} \label{table:ibKAGRA}
    \begin{tabular}{lccc}
      \br
     & iKAGRA & bKAGRA Phase 1 & bKAGRA \\
\hline
Operation year & Mar-Apr 2016 & Apr-May 2018 & 2019- (planned) \\
Interferometer configuration & 3-km Michelson & 3-km Michelson & 3-km RSE \\
Test mass substrate & Fused silica & Sapphire & Sapphire \\
Test mass temperature & room temp. & 18~K/room temp. & 22~K \\
Test mass suspension & Type-C & Type-A & Type-A \\
      \br
    \end{tabular}
  \end{center}
\end{table}

We start this article by describing the full configuration of bKAGRA in Section 2.  
We then describe the details of bKAGRA Phase 1 operation and the results in Section 3. 
Finally, we summarize the current status and our future prospects in Section 4.

%--------------------------------------------------------------------------
% Section 2 
%--------------------------------------------------------------------------

\section{Interferometer configuration of bKAGRA}
\begin{figure}[t]
	\begin{center}
		\includegraphics[width=14cm]{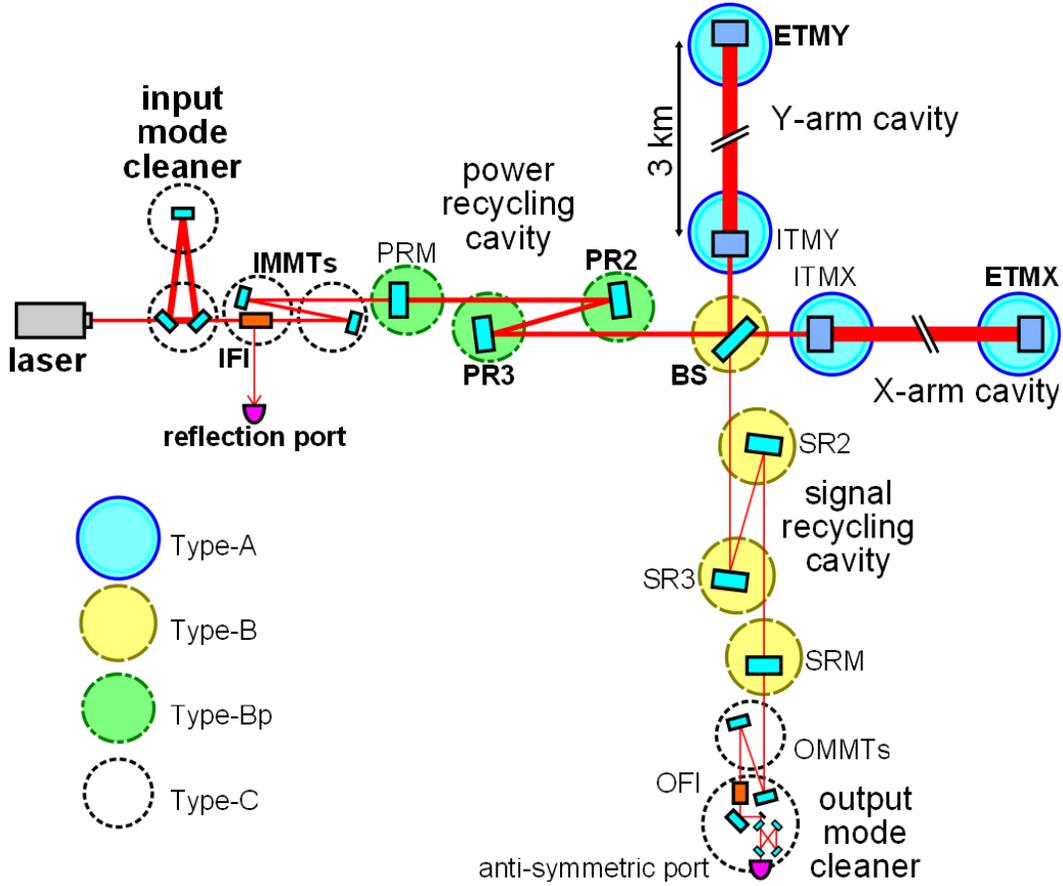}
		\caption{Schematic of the bKAGRA interferometer. All the mirrors shown are suspended inside the vacuum tanks with four types of vibration isolation systems. IMMT (OMMT): input (output) mode-matching telescope, IFI (OFI): input (output) Faraday isolator. Optics used in the Phase 1 operation are labeled in bold.}
		\label{fig:IFOConfig}
	\end{center}
\end{figure}

The schematic of the interferometer of bKAGRA is shown in Fig.~\ref{fig:IFOConfig}. We use a single-frequency continuous-wave laser at a wavelength of 1064~nm. The main part of the interferometer is an RSE  interferometer with two Fabry-P{\'e}rot arm cavities formed by input test masses (ITMs) and end test masses (ETMs) at 22~K~\cite{HiroseSapphire}. Unlike the LIGO and Virgo detectors, we have chosen sapphire as the test mass material because of its high thermal conductivity and high Q value at cryogenic temperatures, which helps in both cooling and lowering thermal noise. We use 22.8~kg ITMs and ETMs whose diameter and thickness are 22~cm and 15~cm, respectively. The size is currently the largest in the market available for high-quality c-axis windows. The system design requires less than 100~ppm round trip loss in the cavities and low-absorption sapphire crystals to reach the target cryogenic temperature. Fabrication of such test mass mirrors was not straightforward and took time, but we have successfully obtained sapphire test mass mirrors which meet the requirement for bKAGRA. The details of the sapphire test masses will be explained elsewhere.

Other mirrors such as the beam splitter (BS), power recycling mirrors (PRM, PR2, PR3), signal recycling mirrors (SRM, SR2, SR3), and input/output mode matching telescope (IMMT/OMMT) mirrors are fused silica mirrors left at room temperature. The substrates of PR2, PR3, SR2 were formerly initial LIGO 25~cm-diameter test masses, and we resurfaced and coated them to match KAGRA's optical configuration. The substrates for the other room temperature mirrors were manufactured by a glass company in Japan, and polishing and coating have been completed as planned.  Diameter of the beam splitter mirror, power/signal recycling mirrors, and input/output mode matching telescope mirrors are 37~cm, 25~cm, and 10~cm, respectively.

The arm cavities have a length of 3~km and a finesse of 1530. The power recycling technique is used to effectively increase the input power by a factor of 10. The SRM is installed at the dark port to extract the GW signal and thus broaden the detector bandwidth. In order to reduce the power inside the ITM substrate to reduce the heat absorption, KAGRA decided to use higher arm cavity finesse than Advanced LIGO and Advanced Virgo. This results in a narrower arm cavity bandwidth, and therefore employing RSE is more important than other detectors. There is an option to detune the signal recycling cavity length to further optimize the spectral shape of the quantum noise to increase the binary neutron star inspiral range~\cite{SomiyaKAGRA,AsoKAGRA}.

The interferometer is also equipped with input and output mode cleaners (IMC and OMC). The IMC is a triangular ring cavity formed by three suspended mirrors, and has a round-trip length of 53.3~m and a finesse of 540. The IMC length is used as a frequency reference above $\sim$1~Hz for pre-stablization of the laser frequency. The OMC is a bow-tie cavity formed by four mirrors monolithically fixed on a base plate, and has a round-trip length of 1.5~m and a finesse of 780~\cite{KAGRAOMC}. The OMC rejects unwanted higher-order spatial modes and frequency sidebands, and the GW signal is extracted from the DC power of the OMC transmitted beam.

\begin{figure}[t]
	\begin{center}
		\includegraphics[width=14cm]{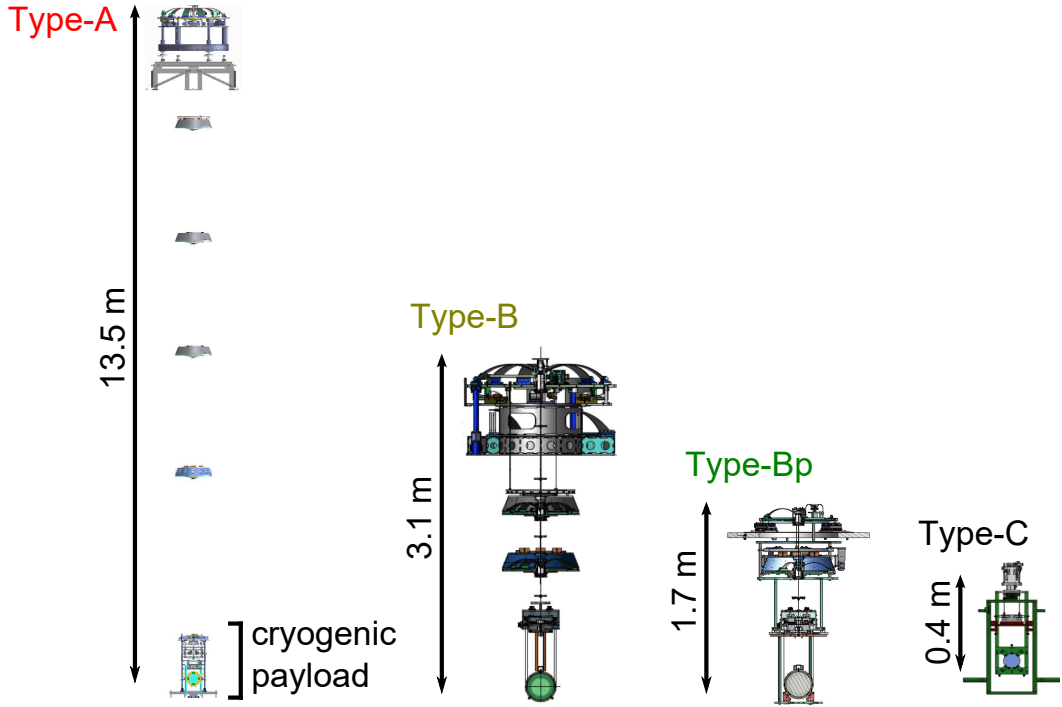}
		\caption{KAGRA mirror suspension system consists of four types. Type-A system is for ITMs and ETMs~\cite{HiroseTypeA, KumarCryopayload}. Type-B system is for BS and signal recycling mirrors~\cite{FabianTypeB}. Type-Bp system is for power recycling mirrors~\cite{ShodaTypeBp}. Type-C system is used for other auxiliary mirrors~\cite{TakahashiTypeC}.}
		\label{fig:SuspensionConfig}
	\end{center}
\end{figure}

\begin{figure}[t]
	\begin{center}
		\includegraphics[width=14cm]{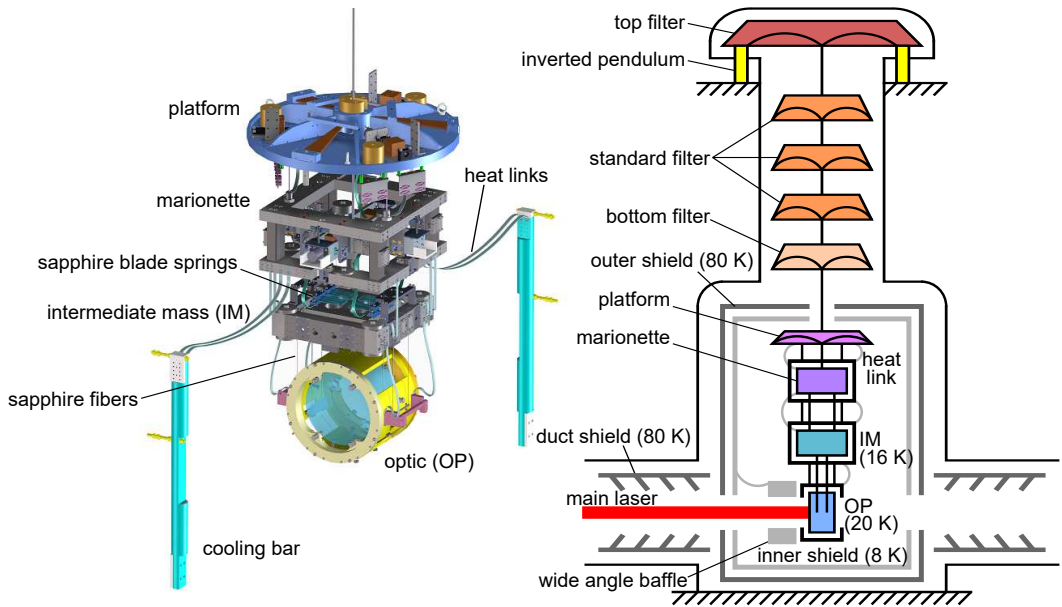}
		\caption{The CAD drawing of the  cryogenic payload under Type-A (left) and the schematic of the cryogenic suspension system of sapphire test masses (right). Suspension stages outside of the outer shield are at room temperature.}
		\label{fig:CryogenicConfig}
	\end{center}
\end{figure}

The interferometer is installed in the vacuum tanks at $10^{-7}$~Pa to mitigate noise from residual gasses. The interferometer mirrors are suspended by four different types of vibration isolation systems inside vacuum tanks, depending on their displacement noise requirements~\cite{MichimuraActuator} (see Fig.~\ref{fig:SuspensionConfig}). ITMs and ETMs are suspended by Type-A system which is an eight-stage pendulum suspended from a top geometric anti-spring (GAS) filter on an inverted pendulum table~\cite{HiroseTypeA}. The GAS filters are for vertical vibration isolation, while the inverted pendulum table is for horizontal vibration isolation. The Type-A system extends over two stories and the legs of the inverted pendulum table are fixed on the second floor. From the top GAS filter, three standard GAS filters, a bottom GAS filter and a cryogenic payload~\cite{KumarCryopayload} are suspended by a single maraging steel fiber in the order mentioned.

The cryogenic payload is the last four stages of the pendulum which are cooled down to cryogenic temperatures. The schematic of the cryogenic suspension system is shown in Fig.~\ref{fig:CryogenicConfig}.  The platform and marionette are suspended from the bottom filter and platform, respectively, with a single maraging steel wire. The intermediate mass is suspended from the marionette with four CuBe fibers. The test mass is in turn suspended from four sapphire blade springs attached to the intermediate mass, with four sapphire fibers (35~cm long, 1.6~mm in diameter). The heat deposited on to the test masses through optical absorption of the arm cavity beam is extracted via these sapphire fibers~\cite{KhalaidovskiSapphireFiber,KomoriThermal}. The intermediate mass is cooled down to about 16~K via high purity 99.9999\% (6N) aluminum heat links attached to the upper stages. The cryogenic payload is surrounded by 8~K inner shields and 80~K outer shields. The cryostat for the test masses also have various kinds of baffles for absorbing stray light inside the cavity, and duct shields at both sides for absorbing the room temperature thermal radiation from vacuum ducts~\cite{SakakibaraShield,AkutsuBaffle}. We use four low vibration double-stage pulse-tube cryocoolers for each cryostat~\cite{SakakibaraCryo,ChenVibration}. The outer shield of the cryostat is cooled down by the first stages of the four cryocooler units. The second stages of two cryocooler units cool the cryogenic payload and the other two cool the inner shield. We also use two single-stage cryocooler units to cool down the duct shields.

For room temperature mirrors, simpler vibration isolation systems are used. BS and signal recycling mirrors are each suspended by a Type-B system which is a four-stage pendulum suspended from a top geometric anti-spring filter on an inverted pendulum table, similar to Type-A~\cite{FabianTypeB}. The power recycling mirrors are each suspended from a triple pendulum called Type-Bp, which is a simplified version of Type-B~\cite{ShodaTypeBp}. Type-Bp system are not supported by an inverted pendulum table. It is instead supported by a set of motorized linear stages, called a traverser, for adjusting the position and the alignment of the suspension chain. The mirrors for the IMC and mode matching telescopes are each suspended from a double pendulum fixed on a three-stage stack for vibration isolation~\cite{TakahashiStack}. This system is called a Type-C system and is a modified version of the suspension used for the TAMA300 GW detector~\cite{TakahashiTypeC}.

\begin{figure}[t]
	\begin{center}
		\includegraphics[width=14cm]{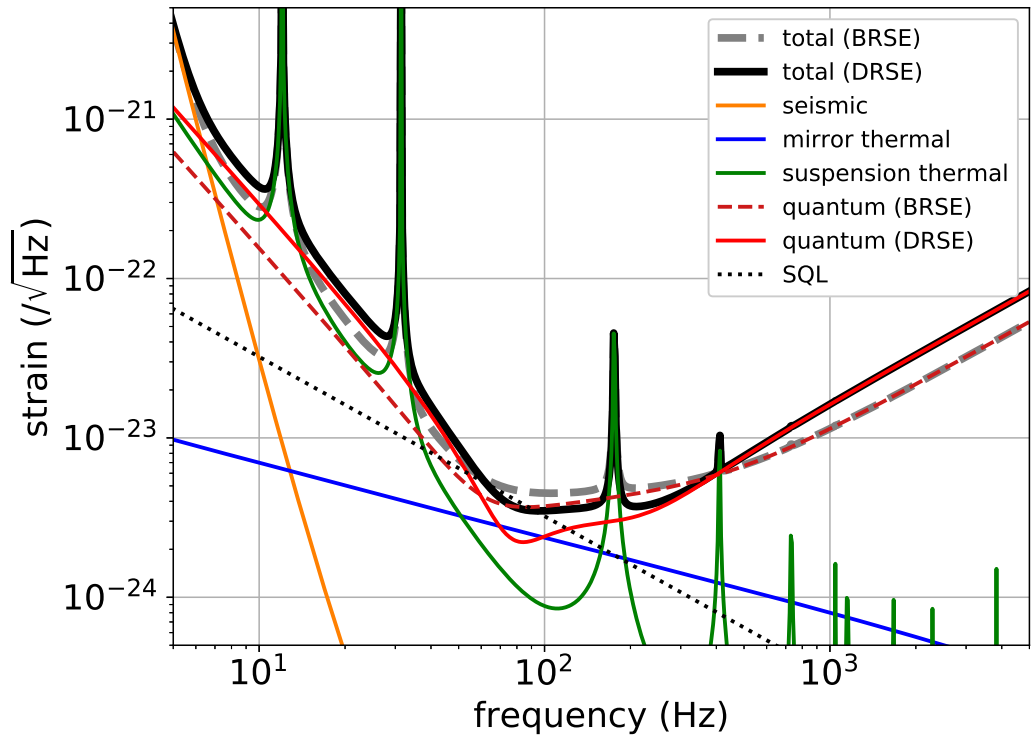}
		\caption{The designed sensitivity of bKAGRA. The quantum noise in both broadband and detuned RSE (BRSE and DRSE) cases are shown. The mirror thermal noise includes terms due to the test mass substrates and the coatings. The seismic noise includes the estimated Newtonian noise from the surface and bulk motion of the mountain containing KAGRA.}
		\label{fig:DesignedSensitivity}
	\end{center}
\end{figure}

Figure ~\ref{fig:DesignedSensitivity} shows the designed sensitivity of bKAGRA. The quantum noise limits the sensitivity for almost the whole frequency band, but the suspension thermal noise and the coating thermal noise also contributes significantly below $\sim$ 100~Hz. To reduce quantum shot noise, higher input power is preferred. However, because of the non-zero absorption of the sapphire mirrors, higher input power results in higher mirror temperature and therefore higher thermal noise. Taking these effects into account, the designed input power at PRM and mirror temperature were chosen to be 67~W and 22~K, respectively, to maximize the binary neutron star inspiral range. The range reaches 135~Mpc in the broadband RSE configuration and 153~Mpc in the detuned RSE configuration with optimal homodyne readout angles. The details of the noise calculations and sensitivity design are described in Refs.~\cite{SomiyaKAGRA,MichimuraPSO}. The main design parameters of bKAGRA is shown in Table~\ref{table:bKAGRAParameters}.

\begin{table}
  \begin{center}
    \caption{The main design parameters of bKAGRA. ITM and ETM have the same radius of curvature and beam radii incident on them. PRC: power recycling cavity, SRC: signal recycling cavity.} \label{table:bKAGRAParameters}
    \begin{tabular}{lc|lc}
      \br
arm cavity length & 3000~m & test mass size & $\phi$22~cm $\times$ 15~cm \\
laser wavelength & 1064~nm & mass of test mass &  22.8~kg \\
input power at PRM & 67~W & test mass temperature & 22~K\\
arm intra-cavity power & 340~kW &  beam radius at test masses & 3.5~cm \\
arm cavity finesse & 1530 & test mass radius of curvature & 1900~m \\
ITM transmittance & 0.4~\% & PRC / SRC lengths & 66.6~m / 66.6~m \\
PRM transmittance & 10~\% & detuning angle & 3.5~deg \\
SRM transmittance & 15~\% & homodyne angle & 135.1~deg \\
      \br
    \end{tabular}
  \end{center}
\end{table}

%--------------------------------------------------------------------------
% Section 3 
%--------------------------------------------------------------------------

\section{bKAGRA Phase 1 operation}
In the Phase 1 operation, two ITMs and optics downstream of the signal recycling cavity were not installed. The PRM was installed but intentionally mis-aligned and we operated the interferometer as a 3-km Michelson interferometer configuration. The optics used in the Phase 1 operation are labeled in bold in Fig.~\ref{fig:IFOConfig}. The GW signal was extracted from the interferometer reflected beam out of the input faraday isolator. We used a heterodyne readout employing modulation-demodulation technique at 16.87293~MHz. The phase modulation depth was measured to be 0.12~rad, and the input power at BS was at around 50~mW. The BS was used as an actuator for the length control of the Michelson interferometer with a control bandwidth of around 50~Hz. The measured interferometer visibility was 99\%. The pressure inside the vacuum tanks varied from $10^{-5}$~Pa to $10^{-4}$~Pa along the 3-km arms.

The ETM for Y arm (ETMY) was suspended by a full cryogenic payload using a Type-A suspension, but the ETM for X arm (ETMX) was suspended by a cryogenic payload without heat links to investigate the effect on the vibration isolation system. The ETMY was cooled down to 18~K, and the ETMX was left at room temperature. One thing to be mentioned here is that a prototype ETMY mirror was used in Phase 1 operation since the high quality mirrors for bKAGRA were not quite ready. The main differences are figure errors in the coated surface and losses in the coating. Although its quality is not as good as the final one, the mirror functioned properly as mentioned in this paper. 

Although the interferometer configuration was very simple, it was the first time that a km-scale interferometer with a sapphire mirror at a cryogenic temperature was operated. Also, it was the first time that all the types of KAGRA suspensions were employed. We also conducted various measurements on mirror installation accuracy, cryogenic cooling, suspensions, and interferometer stability.

\subsection{Initial beam alignment and installation accuracy}
Aligning the laser beam back and forth along the 3-km vacuum tubes requires the mirrors to be aligned to within a few tens of micro-radians. To aid the alignment process, we have installed four photodiodes in front of ETMs, around 10~cm away from the ETM high reflective surface. The four photodiodes are placed concentrically around the center of the ETM so that the readout gives the two-dimensional position of the beam with respect to the center of the ETM. Note that, in final configuration of KAGRA, these photodiodes will be placed on the baffle to capture narrow angle scattering, about 36~m away from the ETM~\cite{AkutsuBaffle}.

We have also installed a telephoto camera (Tcam) system to monitor the position of the beam on the test mass high-reflective surface~\cite{YokozawaTcam}. The Tcam system is similar to the system used to monitor the spot positions of the photon calibrator beams in Advanced LIGO~\cite{aLIGOPCal}. We used a digital camera (Nikon D810) with a telescope (Sky-Watcher BKMAK127) to monitor the ETM surface 36~m away. The long focal distance was necessary to avoid conflict with the cryogenic duct shields and baffles. The camera image was updated at 1~Hz at maximum, and the spatial resolution was around 0.1~mm, which is small enough compared with the size of the beam (3.5~cm in $1/e^2$ radius). The Tcam system was also used to monitor the change in the position of the mirror during the cooling process, and to monitor the spot positions of the photon calibrator beams~\cite{InoueGCal}.

After the alignment of the beams to form a 3-km Michelson interferometer, we have conducted a series of experiments to measure the horizontal and vertical beam positions with respect to the center of the vacuum tubes or vacuum tanks. The discrepancies between the measured and designed positions stayed within 1~cm, which is reasonable considering the installation accuracy of the vacuum tanks. An accurate measurement of the 3-km arm length was not possible in Phase 1 operation, but we have measured the IMC length by measuring the free spectral range. The measured free spectral range was 5.624308(1)~MHz, which corresponds to a round-trip length of 53.30030(1)~m. The difference between the measured and designed length was 1.3~cm.

We therefore concluded that the mirror installation was done at an accuracy of about 1~cm. This is small enough compared with the position and alignment tuning range of the suspensions.

\subsection{Cryogenic cooling of the sapphire mirror}

\begin{figure}[t]
	\begin{center}
		\includegraphics[width=14cm]{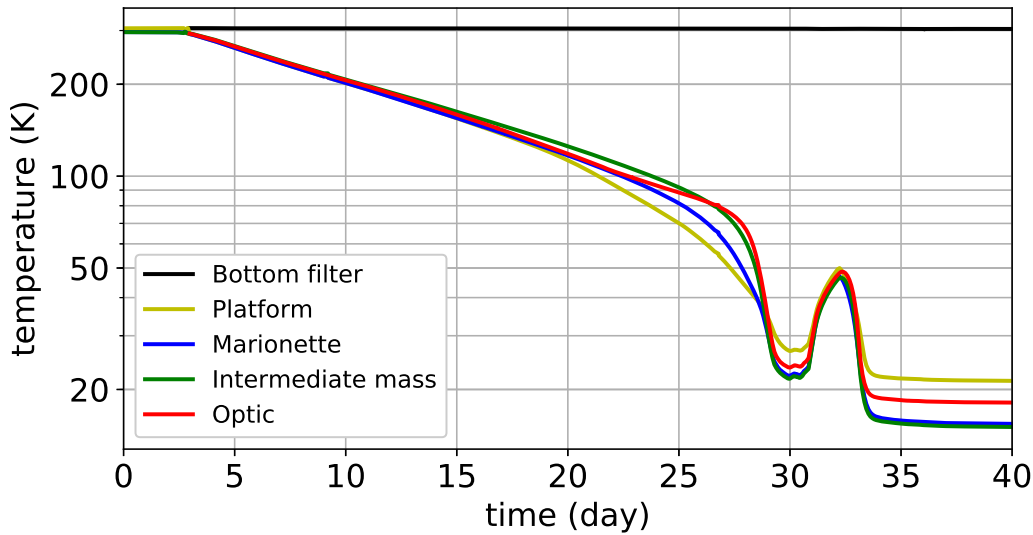}
		\caption{Cooling curve of ETMY. The bump starting from around day 30 was due to the restart of the cryocoolers.}
		\label{fig:CoolingCurve}
	\end{center}
\end{figure}

The cryogenic cooling of ETMY started from the beginning of February 2018 and the temperature of the test mass reached below 20~K in about a month. Fig.~\ref{fig:CoolingCurve} shows the cooling curve of ETMY cryogenic payload for different suspension stages. Until around the 20th day, radiative cooling dominates the cooling process, but after that conduction cooling dominates and the cooling rate is increased~\cite{SakakibaraCryo}. Before reaching 20~K, one of the cryocoolers which cools down the cryogenic payload entered a quasi-stable state at 23~K due to an unbalance of the temperature distribution inside. We therefore stopped and restarted the four cyocoolers. Finally, the sapphire mirror reached 18~K and the intermediate mass reached 16~K. The cooling time and the temperatures reached are almost as expected. Details of the cryogenic system will be given in a separate paper~\cite{UshibaCRY}.

During the cool down, the alignment of ETMY drifted hundreds of micro-radians, mainly in the yaw axis. Therefore, keeping the alignment of the Michelson interferometer continuously during the cool down was not possible. The restoration of the alignment was possible using the alignment actuators on the bottom filter and the marionette stages (see Fig.~\ref{fig:CryogenicConfig}). These alignment actuators have a range of $\pm 18$~mrad for yaw, and $\pm 24$~mrad for pitch. The height change of the test mass due to shrinking of the sapphire fibers and the spring constant change in the blade springs from the cooling is estimated to be around 1~mm and 0.1~mm, respectively. Since pitch drift was around 100~$\mu$rad, we confirmed that variation in the shrinking between different fibers were less than 1\%.

\subsection{Characterization of suspensions} \label{sec:suspensions}
\begin{figure}[t]
	\begin{center}
\begin{minipage}[b]{0.49\textwidth}
   \begin{center}
   \includegraphics[width=8cm]{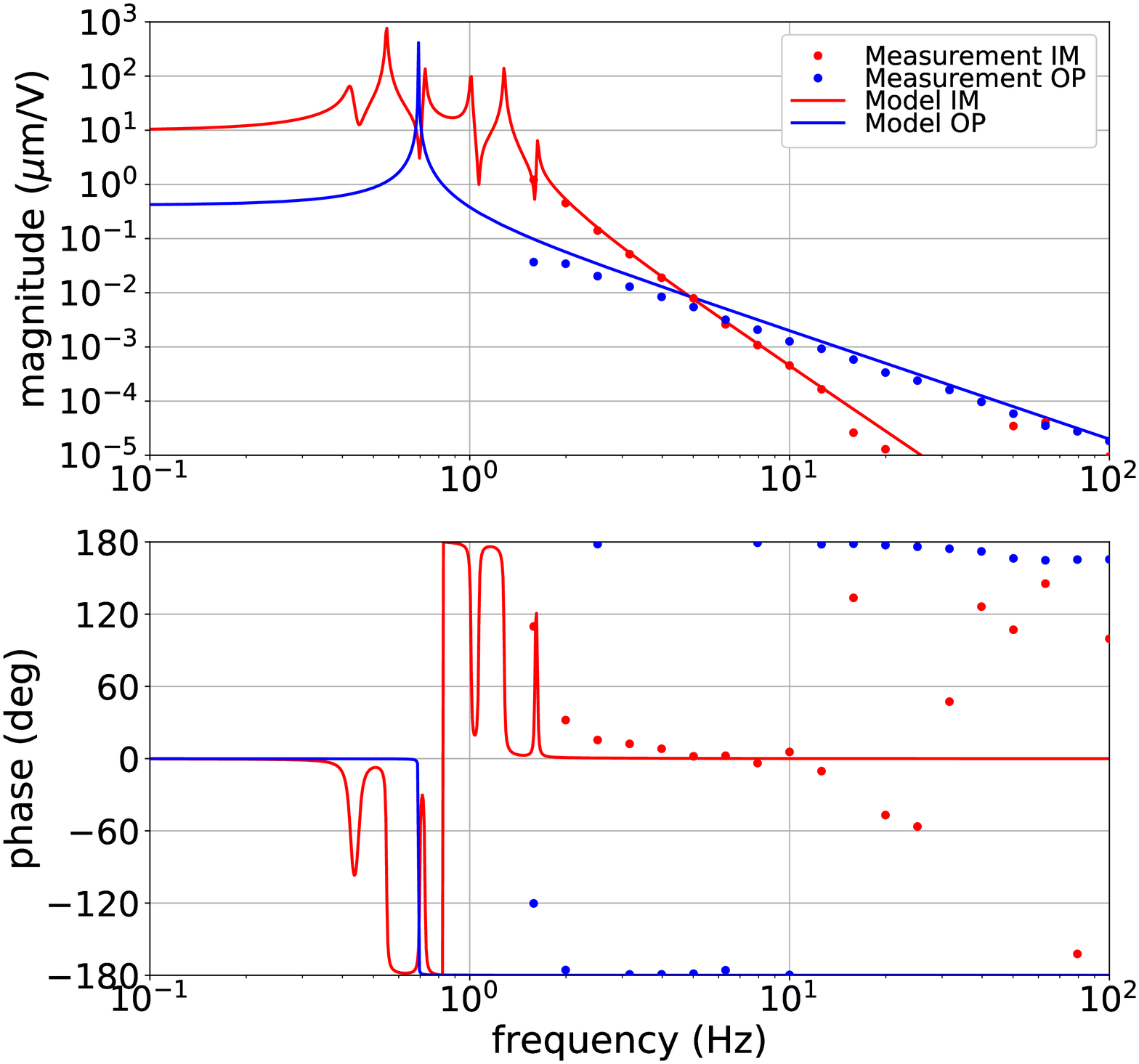} \\
   (a) BS
   \includegraphics[width=8cm]{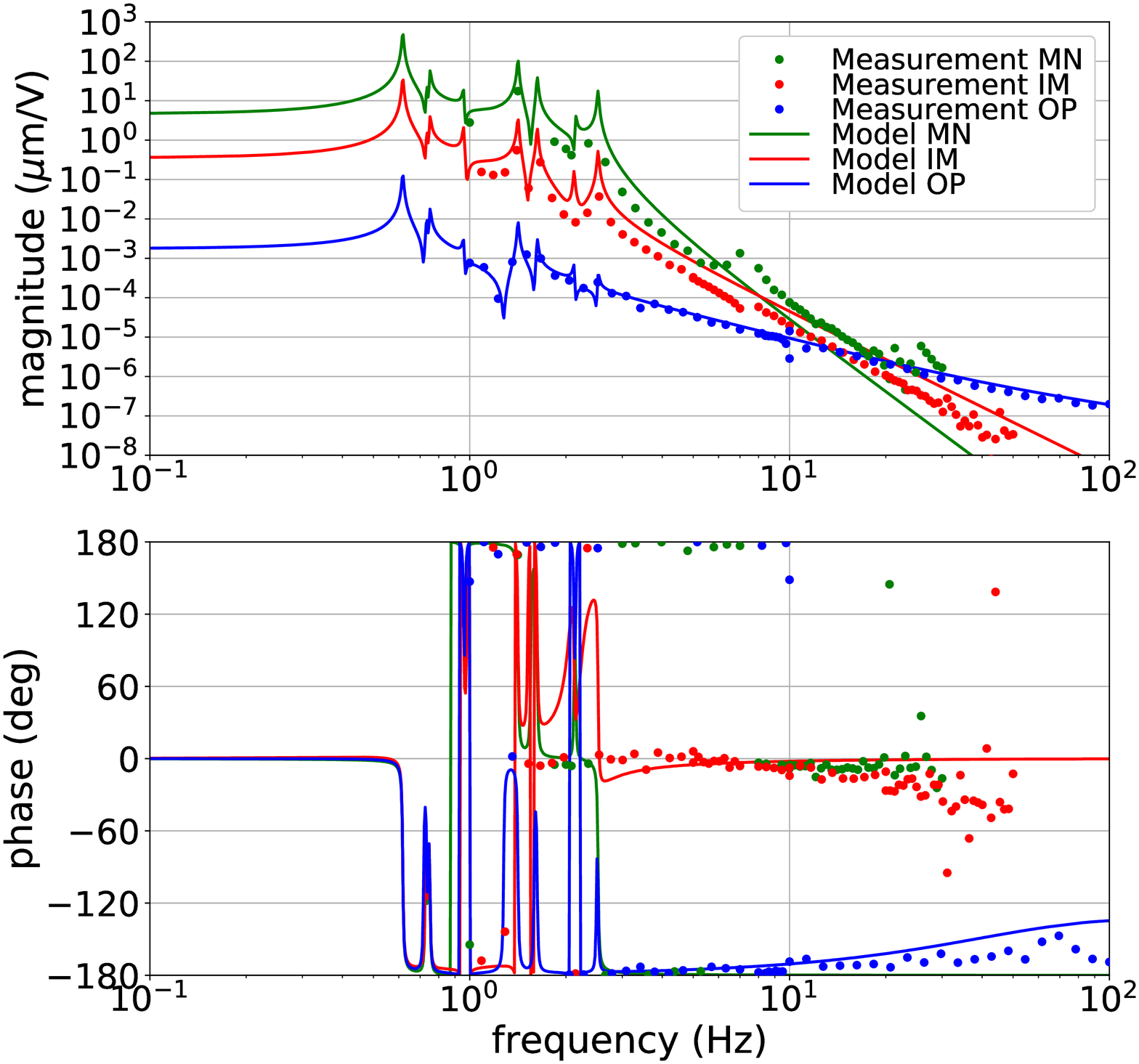} \\
   (b) ETMX
   \end{center}
\end{minipage}
\begin{minipage}[b]{0.49\textwidth}
   \begin{center}
   \includegraphics[width=8cm]{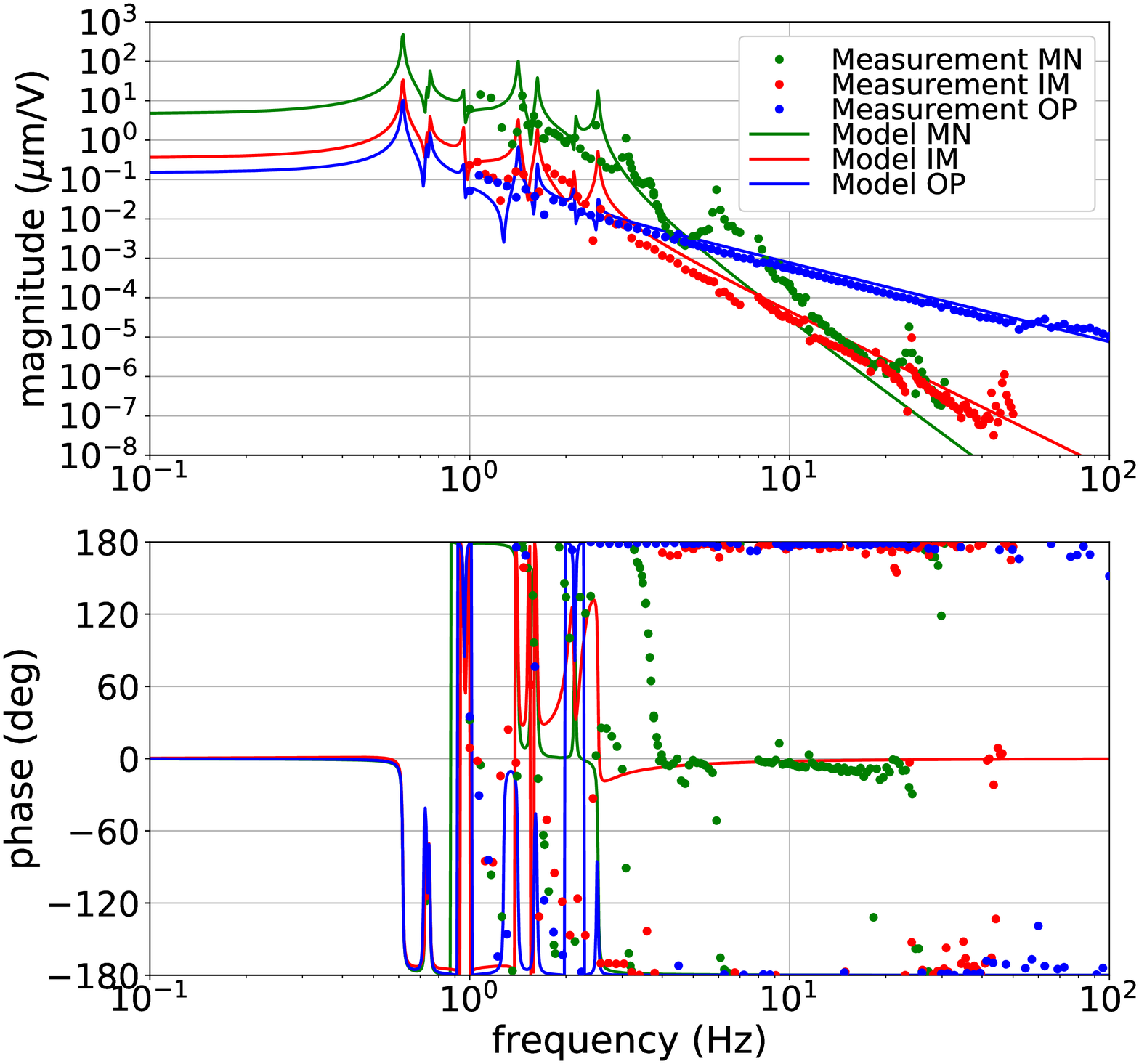} \\
   (c) ETMY
   \end{center}
\end{minipage}
		\caption{Actuation transfer functions of BS (top left), ETMX (bottom left) and ETMY (bottom right). The transfer functions from the voltage applied to the coil drivers to the optic displacements are shown. The measurement results and expected curves from suspension models described in Ref.~\cite{MichimuraActuator} are plotted for marionette (MN), intermediate mass (IM) and optic (OP) stages.}
		\label{fig:ActuationTFs}
	\end{center}
\end{figure}

To operate interferometric GW detectors with high sensitivity, the relative positions and alignments of the mirrors must be finely tuned to maintain the interference conditions. For this purpose, mirror suspension systems are equipped with actuators at various suspension stages. In KAGRA, actuators used for the interferometer length control and alignment control consist of coils and magnets, and actuation is done by controlling the current applied to the coils. Details of the mirror actuation design can be found in Ref.~\cite{MichimuraActuator}.

The performance of the mirror actuation system, especially its performance under cryogenic temperatures, is therefore a critical part of the KAGRA interferometer. During Phase 1, we measured the transfer functions from voltage applied to the coil drivers to the displacement of the mirror using the Michelson interferometer as a displacement sensor, to check the performance. Fig.~\ref{fig:ActuationTFs} shows the result of the measurement, compared with expected curves from suspension models described in Ref.~\cite{MichimuraActuator}. In the suspension model, the transfer functions from the actuator force applied to each suspension stage to the displacement of the mirror are calculated using a suspension rigid-body simulation tool, called SUMCON~\cite{SUMCON}. The coil-magnet actuation efficiencies in the unit of N/A used in the model are the values measured as a standalone actuator.

The measured transfer function matched with the model within a factor of two for BS and ETMX at room temperature. For ETMY at cryogenic temperatures, spurious couplings, especially in the marionette was found. These spurious couplings could be caused by electronics cables from the cryogenic payload to upper stages touching the walls of the vacuum tank. We also found that three standard GAS filters for ETMY were not working properly since some of the parts were mechanically touching the frame. ETMY was the first full Type-A suspension system installed. An improved mechanism for the height adjustment and the installation procedure were employed in ETMX suspension to avoid the same issues. We did not find such issues for ETMX.

Comparison of the ETMX and ETMY actuator transfer function measurements gives the coil-magnet efficiency difference between room and cryogenic temperatures. The coil-magnet efficiency at 20~K calculated from the ETMY transfer function measurements was around 30\% higher than that at room temperature, calculated from ETMX measurements.
We note that the coil drivers used for Phase 1 operation was different from the default ones described in Ref.~\cite{MichimuraActuator}. Except for ETMX optics stage which used the low power and low noise coil driver, high power but higher noise coil drivers are used for BS, ETMX and ETMY suspensions to give a larger actuation range.

\subsection{Interferometer control and calibration}
\begin{figure}[t]
	\begin{center}
		\includegraphics[width=14cm]{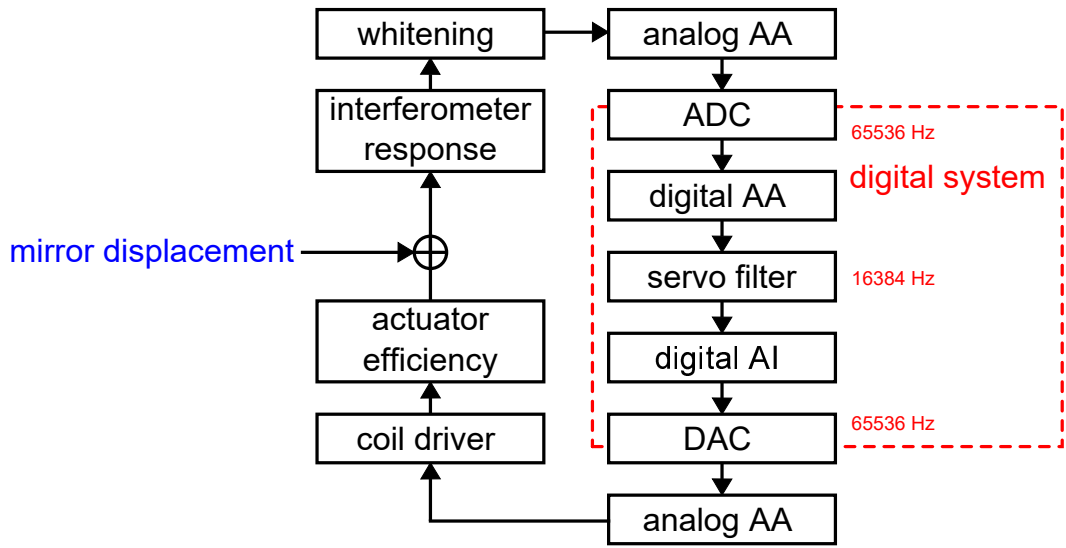}
		\caption{Block diagram of the Michelson interferometer control in Phase 1 operation. AA: anti-aliasing filter, AI: anti-imaging filter.}
		\label{fig:BlockDiagram}
	\end{center}
\end{figure}

The block diagram of the Michelson interferometer control in Phase 1 operation is shown in Fig.~\ref{fig:BlockDiagram}. The differential arm length change caused by mirror displacements is measured with the fringe of the Michelson interferometer. The signal from the photodiode detecting the fringe is sent to the KAGRA digital system via 16-bit analog-to-digital converter (ADC) with a range of $\pm 20$~V. The digital system generates the feedback signal with infinite impulse response servo filters, and the feedback signal are sent to coil drivers via 16-bit digital-to-analog converter (DAC) with a range of $\pm 10$~V. We used both the optic and the intermediate mass stages of BS for the actuation, with a crossover frequency at around 0.4~Hz. Below the cross over frequency, the intermediate mass stage was mainly used. ADCs and DACs operate at a sampling frequency of 65536~Hz and has a total timing delay of 14.6~$\mu$sec. The sampled signals are down-converted to 16384~Hz by the digital system. We used analog 3rd-order Butterworth low-pass filters with a cut-off frequency of 10~kHz for anti-aliasing and anti-imaging. For down-samping and up-sampling in the digital system, we used elliptic low-pass filters. These anti-alialiasing and anti-imaging filters each has a timing delay of 47~$\mu$sec. We used a whitening filter with a zero at 1~Hz and a pole at 10~Hz to effectively reduce the ADC noise.

The calibration of the mirror displacement was done using the fringe of the Michelson interferometer when the mirror is freely swinging. The calibration factor from the mirror displacement to the error signal, which we call optical gain, was typically measured to be $(5.7 \pm 0.7) \times 10^{7}$~V/m. The uncertainty mainly comes from the fluctuation of the interferometer fringe from alignment fluctuations of the mirrors.

\begin{figure}[t]
	\begin{center}
\begin{minipage}[b]{0.49\textwidth}
   \begin{center}
   \includegraphics[width=7cm]{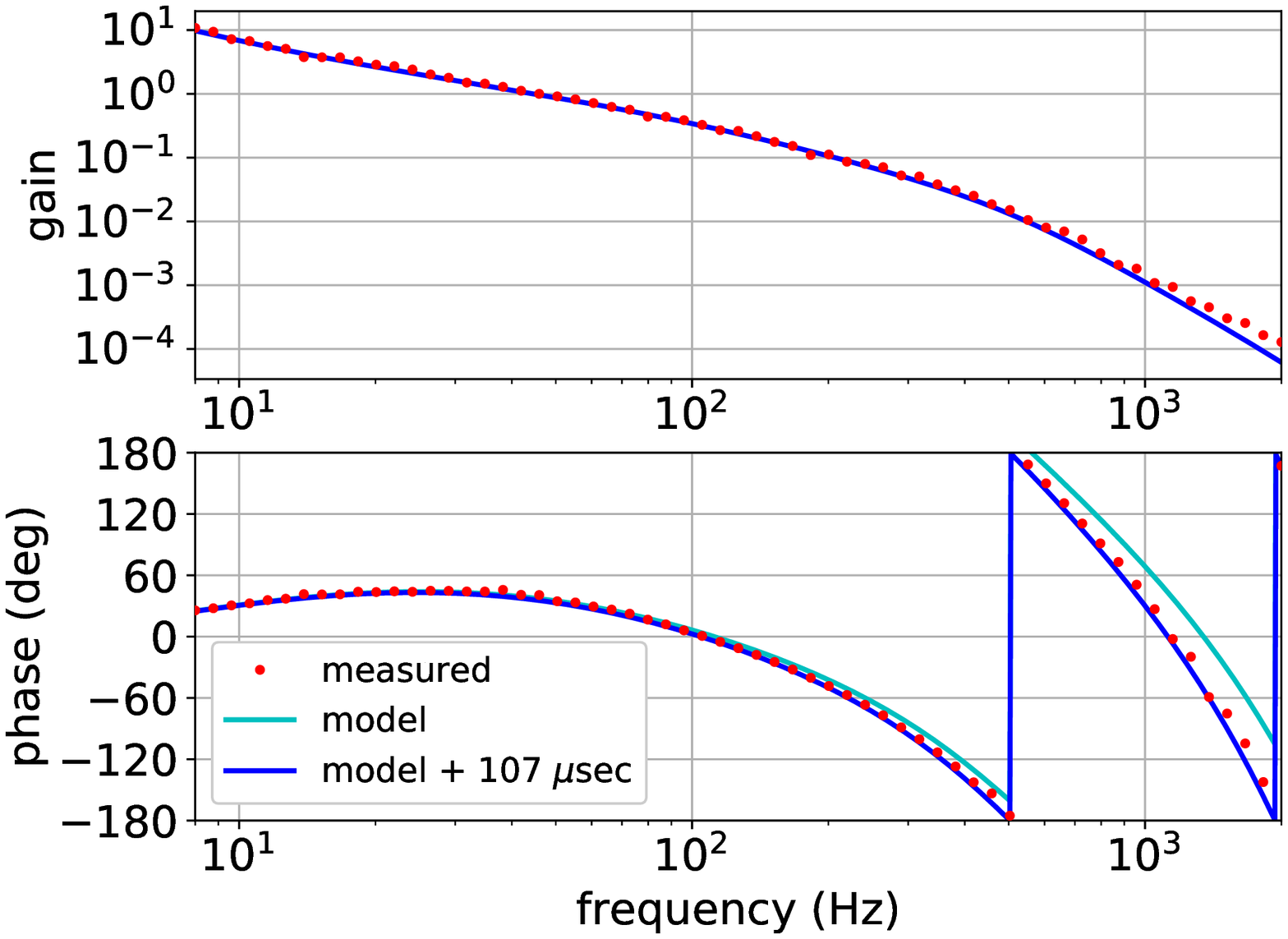} \\
   \end{center}
\end{minipage}
\begin{minipage}[b]{0.49\textwidth}
   \begin{center}
   \includegraphics[width=7cm]{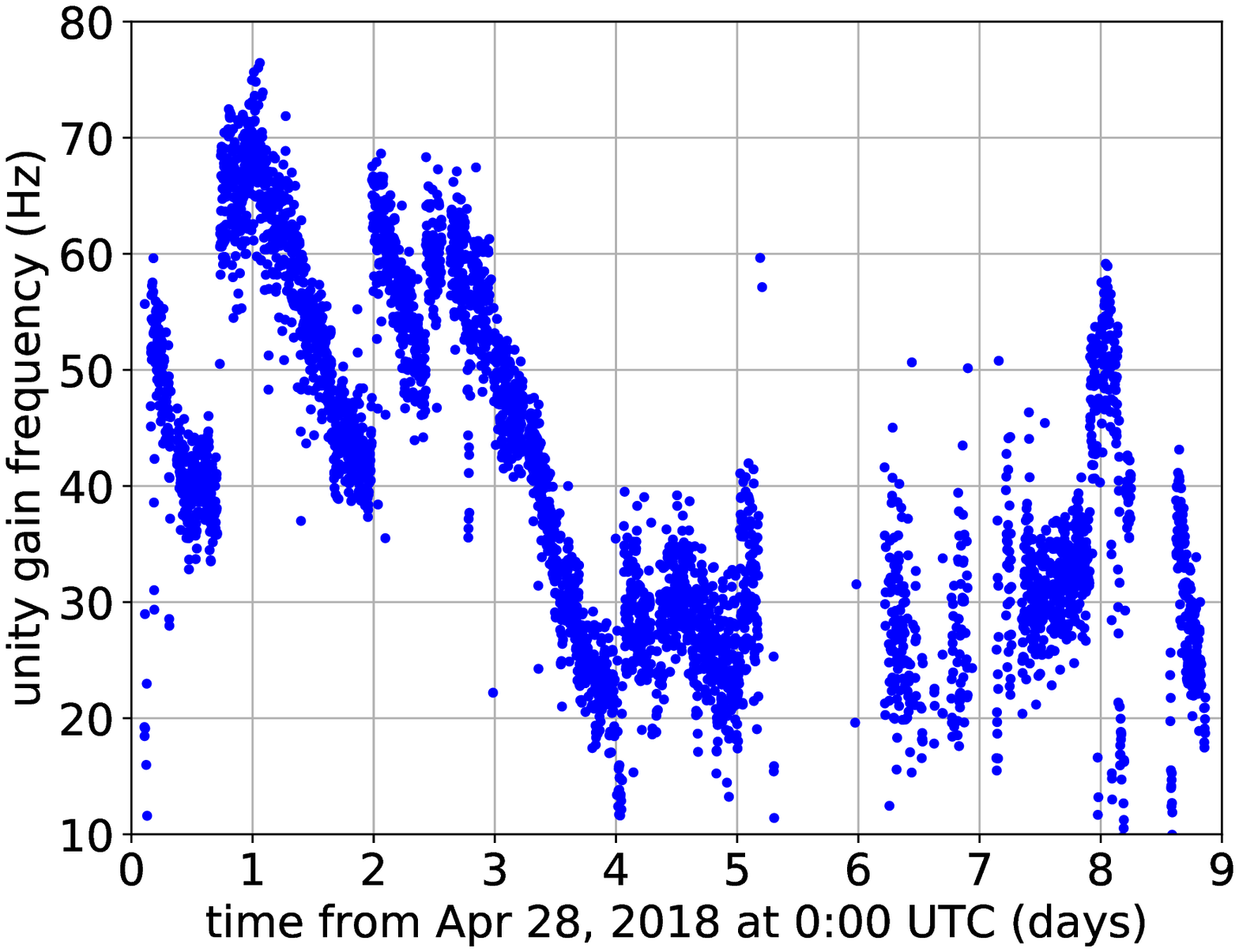} \\
   \end{center}
\end{minipage}
		\caption{The openloop transfer function of the Michelson interferometer control (left) and estimated unity gain frequency over time (right). The expected openloop transfer function matches with the measurement by adding 107~$\mu$sec extra phase delay.}
		\label{fig:OpenloopTF}
	\end{center}
\end{figure}

The measured openloop transfer function of the servo is shown left in Fig.~\ref{fig:OpenloopTF}. The unity gain frequency of the servo was typically at 50~Hz. To monitor the openloop transfer function change over time, we have injected three calibration lines, at 15~Hz, 51~Hz, and 91~Hz, using BS. The unity gain frequency drift calculated from the calibration line at 91~Hz is shown right in Fig.~\ref{fig:OpenloopTF}. The overall gain drifted by about 50\% per an hour.
Although it is generally not possible to distinguish what is the cause of this drift, we have occasionally measured the optical gain and actuator efficiency during the operation when the interferometer is not locked, and concluded that it is likely to be from the optical gain drift. The optical gain drift was caused by visibility degradation from the Michelson interferometer alignment drift. This was also the case in iKAGRA operation.

From the openloop transfer function measurement, we have also measured the overall timing delay of the loop. The measured delay was $520 \pm 3~\mu$sec, and we had an unexpected delay of about $107~\mu$sec. Our model accounts for all the timing delays mentioned above and the time it takes for transferring signals between separated digital control computers (168~$\mu$sec), giving a total of $413~\mu$sec.

The input and output of the servo filter in the digital system are called the error signal $v_{\rm err}$ and the feedback signal $v_{\rm fb}$, respectively. Using these signals, we have reconstructed the Michelson differential arm length change in the frequency domain with
\begin{equation}
 \delta L = \frac{1}{C} v_{\rm err} + A v_{\rm fb},
\end{equation}
where $C$ is the optical gain and $A$ is the actuation efficiency. The reconstruction was done online using infinite impulse response filters, and the latency was less than 1 sec.

\subsection{Data transfer}
\begin{figure}[t]
	\begin{center}
		\includegraphics[width=14cm]{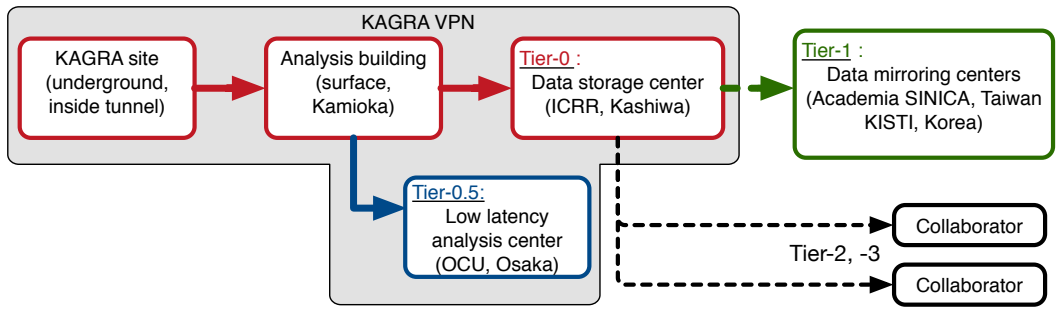}
		\caption{Overview of the KAGRA data transfer system. Tier-0 (original) and Tier-0.5 (low latency) are formed as isolated local network using VPN (Virtual Private Network).}
		\label{fig:DataTransfer}
	\end{center}
\end{figure}

The acquired KAGRA interferometer signals, some channels inside the digital control servo loop and other monitor channels are recorded into frame format files in 32 second chunks and are transferred to multiple data centers as shown in Fig.~\ref{fig:DataTransfer}. 
The frame format is an international common data format of GW  experiments~\cite{frameformat}.
All the raw data from the KAGRA detector inside the tunnel is first transferred to a control room in the analysis building on the ground via 4.5-km optical fiber links. The total data rate is expected to reach 20~MB/s in the final phase of bKAGRA. The analysis building has a 200~TB data storage system which can store about four months of data. The raw data and processed data including calibrated strain signal $h(t)$ are then transferred to the data storage center at the Institute for Cosmic Ray Research in Kashiwa and the low latency data analysis center at Osaka City University. The data storage center has a storage system with a capacity of 2.5~PB. The low latency data analysis center has 760~CPU cores.
We also transfer the data to data mirroring centers in Taiwan and Korea. 
The strain signal data can be exchanged with other GW detectors such as LIGO and Virgo in future observational operations.

During Phase 1, we also tested the data transfer system for various use cases: analysis of hardware injected waveforms, development of event search pipe-lines, detector characterization, and test of the data transfer system itself, etc. 
The measured data transfer speed from the KAGRA detector to the analysis building was about 200~MB/s, which exceeds the requirement of 20~MB/s. The measured latencies of the data transfer from the KAGRA site to the low latency analysis site in Osaka was about 3~seconds. 

We have been developing data analysis tools at the data storage center in Kashiwa.
Daily spectra of the major channels are generated automatically and served via a web page to the collaborators.
To realize very low latency off-line searches, data from the frame file is transmitted to multiple servers with shared memory dedicated to these calculations.
We have also implemented a matched filtering analysis for compact binary coalescences.

\subsection{Interferometer sensitivity and stability}

\begin{figure}[t]
	\begin{center}
		\includegraphics[width=14cm]{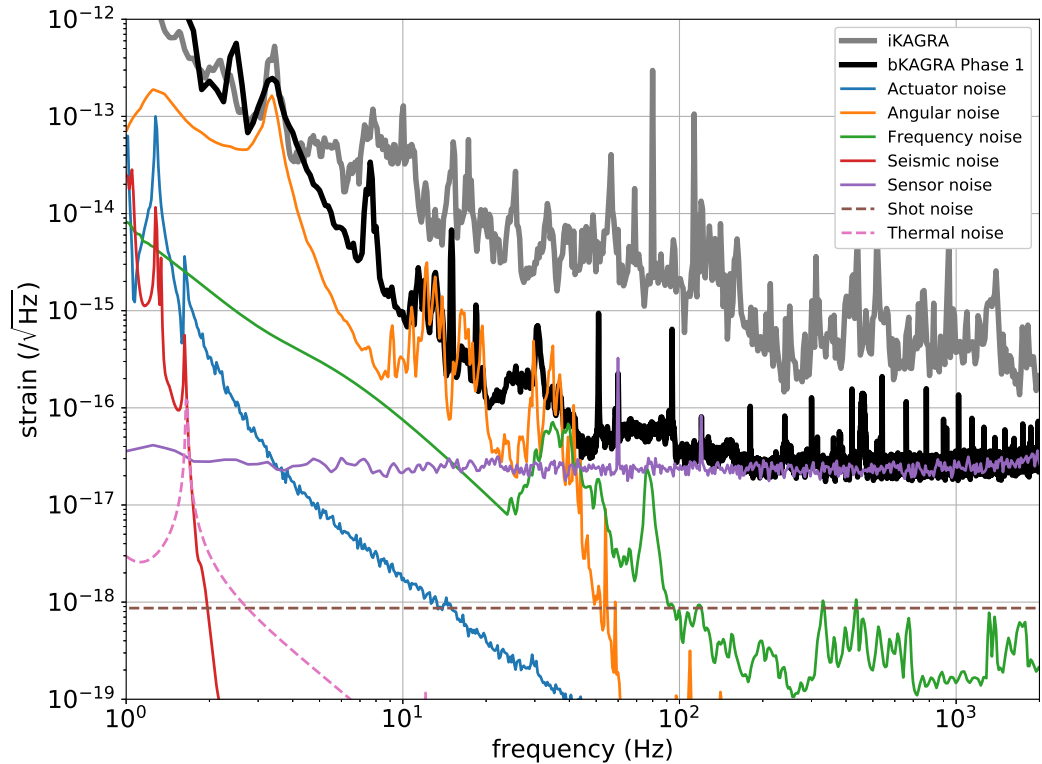}
		\caption{The strain sensitivity of KAGRA during Phase 1 operation, compared with various strain equivalent noises. Actuator noise is the sum of the displacement noise of the mirrors from electronics noise for the actuation~\cite{MichimuraActuator}. Angular noise is the sum of the mirror angular fluctuation coupled to the displacement signal. Frequency noise is the estimated laser frequency noise suppressed through laser frequency stabilization using IMC. Seismic noise is estimated mirror displacement noise from the ground displacement attenuated by the mirror suspensions. Sensor noise is the sum of ADC noise and the dark noise of the photodiode. Thermal noise is the estimated suspension thermal noise of ETMX. That for ETMY is smaller by more than an order of magnitude due to cooling. The peaks at 80~Hz and 113~Hz for iKAGRA, and those at 15~Hz, 51~Hz and 91~Hz for Phase 1 are from the calibration lines.}
		\label{fig:NoiseBudget}
	\end{center}
\end{figure}

Although Phase 1 was not meant for low noise operation to detect GWs, we have also checked the strain sensitivity of the detector. A typical sensitivity curve during the Phase 1 operation is shown in Fig.~\ref{fig:NoiseBudget}. Below around 50~Hz, the sensitivity was limited by angular control noise, mainly from BS motion in yaw. During Phase 1, angular motions of PR2, BS, ETMX and ETMY are controlled using optical levers as angular sensors~\cite{KokeyamaTiltMeter}. Above 100~Hz, the sensitivity was limited by the dark noise of the photodiodes receiving an error signal for the Michelson interferometer length control. The dark noise was quite high since the photodiodes used are optimized for the full power operation of bKAGRA. Compared with iKAGRA, sensitivity was improved by more than an order of magnitude mainly due to reduced acoustic noise coupling. In iKAGRA, acoustic noise was high because vacuum chambers were not evacuated~\cite{iKAGRA}.

\begin{figure}[t]
	\begin{center}
		\includegraphics[width=14cm]{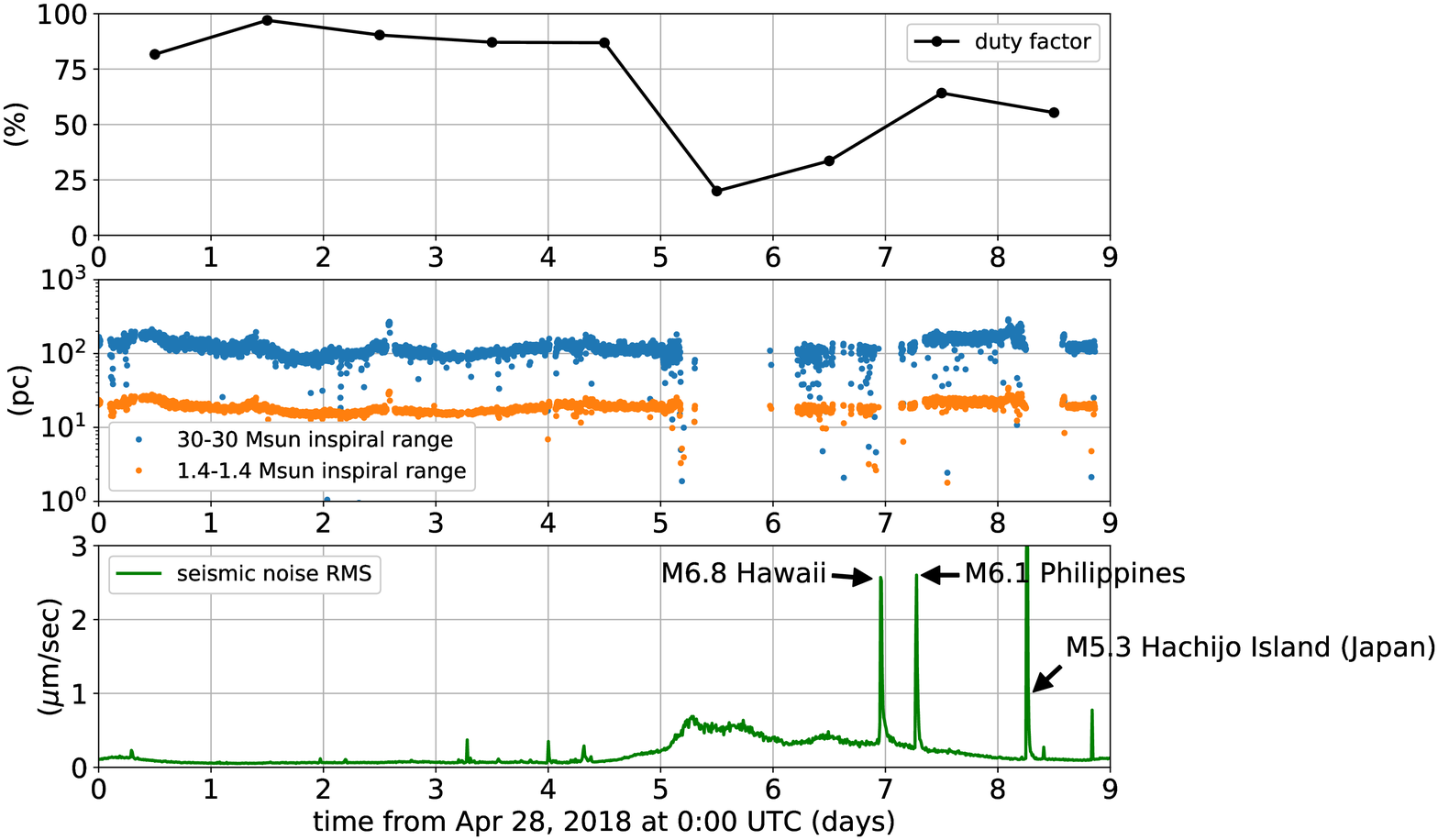}
		\caption{Daily duty factor (top), inspiral range (middle), and seismic noise level (bottom) during Phase 1 operation. The root-mean-square (RMS) of the ground velocity measured with a seismometer is shown.}
		\label{fig:Stability}
	\end{center}
\end{figure}

Fig.~\ref{fig:Stability} shows the daily duty factor, inspiral range, and root-mean-square of the seismometer output. The inspiral range for binary neutron stars was about 17~pc and that for $30M_{\odot}-30M_{\odot}$ binary black holes was about 100~pc. The overall duty factor, the ratio of the locked period to the whole operation period, was 69\%. The micro-seismic noise was higher at the last half of the operation due to local stormy weather, and the interferometer was not stable. Also, big earthquakes in Hawaii, Philippines, and Hachijo Island of Japan affected the interferometer. According to the United States Geological Survey, their magnitude was 6.8, 6.1 and 5.3, respectively. The effect of seismic motion was higher than expected due to rubbing issues in the ETMY suspension, as discussed in Section~\ref{sec:suspensions}.

The duty factor during the first half of the operation when the seismic noise was lower reached 88.6~\%.
The lock of the interferometer was done automatically using a state machine automaton called {\it Guardian}~\cite{Guardian}. The recovery of the lock often took less than 10~minutes from lock loss. The longest stretch in which lock was maintained was 11.1~hours, which was achieved in May 3.

%--------------------------------------------------------------------------
% Section 4
%--------------------------------------------------------------------------
\section{Conclusions and outlook}

We have operated a 3-km Michelson interferometer with a cryogenic sapphire mirror for the first time. The sapphire mirror was suspended by a full eight-stage pendulum, and was successfully cooled down to below 20~K within 35 days. The alignment drift during the cool down and mirror actuation efficiency were measured, and we confirmed that interferometer alignment and control is possible at cryogenic temperatures. We have also checked the mirror installation accuracy, and confirmed that the accuracy is good enough to form a full interferometer. There were some issues regarding rubbing of electric cables and stuck GAS filters in the first test mass suspension system installed, but these issues were successfully resolved in the second system.

The Phase 1 operation marked the first step in realizing a full 3-km cryogenic interferometer at an underground site. 
After the Phase 1 operation, we have been installing more mirror suspension sytems and a higher power laser source to form a full RSE interferometer at cryogenic temperatures.  We expect to finish all installation work by the end of March 2019, and start commissioning of the detector then.  The first scientific observation run shall be started in late 2019.

%--------------------------------------------------------------------------
% Acknowledgement and References
%--------------------------------------------------------------------------
\ack
We would like to thank the Advanced Technology Center (ATC) of NAOJ, the Mechanical Engineering Center of KEK, and the machine shop of the Institute of Solid State Physics (ISSP) of the University of Tokyo for technical support.

This work was supported by MEXT, JSPS Leading-edge Research Infrastructure Program, JSPS Grant-in-Aid for Specially Promoted Research 26000005, JSPS Grant-in-Aid for Scientific Research on Innovative Areas 2905: JP17H06358, JP17H06361 and JP17H06364, JSPS Core-to-Core Program A. Advanced Research Networks, JSPS Grant-in-Aid for Scientific Research (S) 17H06133, the joint research program of the Institute for Cosmic Ray Research, University of Tokyo, National Research Foundation (NRF) and Computing Infrastructure Project of KISTI-GSDC in Korea, Academia Sinica (AS), AS Grid Center (ASGC) and the Ministry of Science and Technology (MoST) in Taiwan under grants including AS-CDA-105-M06, the LIGO project, and the Virgo project. This paper carries JGW Document Number JGW-P1809289.

\end{document}